\documentstyle[11pt,aaspp4,psfig]{article}   

\def\Msun{\,M$_\odot$}
\def\Rsun{\,R$_\odot$}

\def\PCB{$P_{\rm CB}$}
\def\PWB{$P_{\rm WB}$}
\def\PS{$P_{\rm S}$}

\lefthead{Kalogera \& Lorimer}
\righthead{Upper Limit on NS/NS Coalescence Rate}

\begin{document}

\slugcomment{To appear in The Astrophysical Journal.}

\title{An Upper Limit on the Coalescence Rate of Double Neutron-Star
Binaries in the Galaxy}
\author{Vassiliki Kalogera}
\affil{Harvard-Smithsonian Center for Astrophysics, 60
Garden St., Cambridge, MA 02138; vkalogera@cfa.harvard.edu}
\author{Duncan R. Lorimer}
\affil{Arecibo Observatory, HC3 Box 53995, Arecibo, Puerto Rico, 
PR 00612 ; dunc@naic.edu}

\begin{abstract}

In the context of assessing the detectability of the coalescence of
two neutron stars (NS) by currently built gravitational-wave experiments, 
we present a way of obtaining an upper limit to the coalescence rate in
the Galaxy. We consider the NS/NS progenitors just before the second   
supernova explosion. We combine the theoretical understanding of orbital
dynamics at NS formation with methods of empirically estimating pulsar
birth rates and derive an upper limit of a few mergers every $10^5$\,yr.
Such a Galactic rate implies a possible detection by the ``enhanced'' LIGO
of up to a few to ten mergers per year.

\end{abstract}

\keywords{binaries: close --- stars: neutron --- pulsars: general}

\section{Introduction}

With the upcoming completion of the first-level functional stage of
gravitational-wave detection experiments (LIGO, VIRGO, GEO600) currently
under development, a serious effort has been focused on studying possible
sources of gravitational radiation and on examining their detectability.
The coalescence of two compact objects, neutron stars (NS) and/or black
holes (BH), is one of the primary sources expected to be detected out to
$\sim 20$\,Mpc with the first-generation LIGO and out to $\sim
200$\,Mpc with the ``enhanced'' LIGO (e.g., \cite{A92}1992;
\cite{T96}1996). Apart from the strength of the expected signal, a major
factor in assessing the detectability of such sources is the frequency of
coalescence events in the Galaxy. This Galactic rate provides the basis
for estimates of the detection rate depending on the detector sensitivity.
For the expected sensitivity of ``enhanced'' LIGO, a Galactic NS/NS merger
rate of $10^{-5}$\,yr$^{-1}$ would lead to a detection rate of $2-3$
events per year (using the extrapolation out to 200\,Mpc by
\cite{P91}1991).

So far two ways have been used to estimate the Galactic rate of compact
binary mergers. One is purely theoretical and relies on modeling the
formation of such systems through a long series of evolutionary stages by
means of population synthesis. This method has been used by several
different groups to obtain coalescence rate estimates for NS/NS, BH/NS
and BH/BH binaries in recent years. However, the calculations involve a
number of uncertain factors related to the properties of primordial
binaries and to the details of stellar and binary evolution. Detailed
parameter studies to examine the robustness of the results is a crucial
component of the population synthesis calculations. The results on the
binary compact object merger rate in our Galaxy from a number of recent
studies (\cite{F99b}1999; \cite{B99}1999; \cite{B98}1998;
\cite{PZ98}1998; \cite{F98}1998; \cite{L97}1997) cover a range of values
of $\sim 3-4$ orders of magnitude ($\lesssim 10^{-7}$ to $\gtrsim
10^{-4}$\,yr$^{-1}$). The major uncertainty turns out to be the average
magnitude of kicks imparted to newborn compact objects. For kicks in a
range from very low values to several hundreds of km\,s$^{-1}$, the
coalescence rate typically varies by more than two orders of magnitude.  
Other uncertainties include the physical assumptions made when modeling
common envelope evolution (particularly the fate of NS), the mass-ratio
distribution in primordial binaries, and the progenitor masses for BH.
Overall it appears that the predictive power of population synthesis
calculations particularly for the coalescence rate of NS/NS, BH/NS, and
BH/BH binaries is rather limited (see \cite{K99}1999). 

For binaries that are known to exist in our Galaxy, i.e., coalescing NS/NS
binaries, there is also an empirical way of estimating their merger rate.
It is based on modeling radio pulsar selection effects (for the pulsar
surveys conducted so far). The results of such studies (\cite{N91}1991;
\cite{P91}1991; \cite{C95}1995; \cite{A98}1998; \cite{E99}1999)  appear to
cover a much narrower range (about an order of magnitude wide) ranging
roughly from $\sim 10^{-7}$ to $\sim 10^{-6}$\,yr$^{-1}$. For a more
detailed discussion of the uncertainties involved see \cite{A98}(1998) and
\cite{KN99}(1999).

Given on the one hand all the uncertainties of the various rate
estimates and on the other the significance of a reasonable detection
rate for the gravity wave experiments, attempts to obtain a secure
{\em upper} limit on the coalescence rate have been made. Specifically for
NS/NS binaries, \cite{B96}(1996) derived an empirical upper limit based on
the non-detection of a young (non-recycled) pulsar among the observed
NS/NS binaries. He used the total number of young radio pulsars detected
at the time and their birth rate and concluded that the NS/NS formation
rate cannot exceed $10^{-5}$\,yr$^{-1}$. More recently, \cite{A98}(1998)
revised this upper limit taking into account current pulsar numbers and
the reduction in search sensitivity to short orbital period binary systems
due to Doppler shifting of the pulse period during an observation. These 
considerations led Arzoumanian et al.~to increase Bailes' upper limit to 
$10^{-4}$\,yr$^{-1}$. 

In this paper we present an alternative way of obtaining an upper limit to
the NS/NS coalescence rate in our Galaxy that combines our basic
theoretical understanding of the late stages in the evolutionary sequence
of NS/NS formation with empirical birth rate estimates of other radio
pulsar systems that contain pulsars similar to the ones detected in NS/NS
binaries. In the next section we describe the basic method for deriving
the upper limit. In \S\,3 we deal quantitatively with the problem of
supernova orbital dynamics and the formation of the second NS in NS/NS
binaries. In \S\,4 we obtain an empirical estimate of the birth rate of
single pulsars similar to those found in NS/NS binaries and in \S\,5 we
calculate the upper limit on the NS/NS coalescence rate in our Galaxy and
compare it to the requirements for gravitational-wave detection. Finally,
in \S\,6 we discuss in detail some of the uncertainties related to the
derived upper limit as well as the prospects of constraining this limit
further in the future.

\section{Basic Method}

We consider the {\em immediate} progenitors of NS/NS binaries, i.e., the
stage just before the second supernova (SN) explosion. Regardless of the
details of prior evolution all different formation mechanisms discussed in
the literature (e.g., \cite{B91}1991; \cite{B95}1995; \cite{T95}1995)
converge to the same binary configuration prior to the formation of the
second NS: a binary consisting of the first-born NS and a helium star
(He-star) in a circular orbit. The implicit assumption in these formation
channels is that the observed radio pulsar is the first-born NS in the
binary and the orbit before the second explosion has been circularized
through a mass-accretion phase, during which the NS also got (mildly)
recycled.

Depending on its mass ($\gtrsim 2-3$\Msun \,and $\lesssim 10$\Msun;
\cite{H85}1985; \cite{F99a}1999), the He-star can terminate its life in a
supernova explosion and the formation of a NS. During the explosion the
binary characteristics change not only because of mass loss but also
because of a birth kick imparted to the compact remnant. Although the
physical origin of this kick is not well understood, its existence is
supported by the characteristics of a number of different NS populations
(isolated and in binaries; e.g., \cite{vdH97b}1997; see however
\cite{I98}1998). There are three possible outcomes after this SN
explosion: (i) a NS/NS binary is formed in a tight enough orbit so that
the system coalesces due to gravitational radiation within $10^{10}$\,yr;
coalescing binary (CB), (ii) a NS/NS binary is formed but the orbit is
too wide for coalescence to occur within $10^{10}$\,yr; wide binary (WB),
and (iii) the binary gets disrupted and a single (S), mildly recycled
pulsar is freed. The characteristics discriminating between the three
results are the post-SN orbital semi-major axis, $A$, and eccentricity,
$e$. In Figure 1, the three different types of systems are identified on
the $A-e$ plane for coalescence within $10^{10}$\,yr.

For specific values of the mass of the He-star, $M_o$, the pre-SN orbital
separation, $A_o$, and the magnitude, $V_k$, of the kick imparted to the
second NS, and under the assumption of isotropic kicks, we can calculate
the branching probabilities for each of the above outcomes, \PCB , \PWB ,
and \PS . All three subgroups contain pulsars with the same properties as
the ones found in the observed coalescing NS/NS binaries, since the pulsar
was born and recycled prior to the ``immediate-progenitor'' stage we
consider here. Therefore the ratio of two of the above branching
probabilities is equal to the ratio of birth rates for the corresponding
two types of systems. In what follows we choose to consider the coalescing
binaries and the disrupted systems:
 \begin{equation}
 \frac{P_{\rm CB}}{P_{\rm S}}~=~\frac{BR_{\rm CB}}{BR_{\rm S}},
 \end{equation}
 where $BR_{\rm CB}$ and $BR_{\rm S}$ are the birth rates of coalescing
NS/NS and single, mildly recycled pulsars.

To obtain an exact value of the ratio of branching probabilities, $P_{\rm
CB}/P_{\rm S}$, the distribution functions of the three input parameters,
$M_o$, $A_o$, and $V_k$, are needed. This however requires modeling of the
complete binary evolution prior to the stage we consider here, which
brings us back to the problem of all the uncertainties that affect the
population synthesis calculations of NS/NS binaries, because of which the
calculated probability ratio would be uncertain by several orders of
magnitude.  We can avoid all these uncertainties by calculating an {\em
upper bound} to the ratio rather than its actual value. This upper bound,
which requires knowledge of the allowed ranges of values for $M_o$, $A_o$,
and $V_k$ but not their distribution functions, can then be used to derive
an upper limit to $BR_{\rm CB}$. The missing link then is $BR_{\rm S}$, an
estimate of which we obtain empirically (as is usually done for coalescing
NS/NS binaries) based on the observed pulsar sample. Using the two
components, ($P_{\rm CB}/P_{\rm S}$)$^{\rm max}$ and $BR_{\rm S}$, we can
derive an upper limit to the NS/NS coalescence rate in our Galaxy. In the
next two sections we describe our assumptions and the derivation in
detail.

\section{Supernova Orbital Dynamics}

We consider the immediate progenitors of NS/NS binaries consisting of the
first NS (pulsar) of mass $M_{\rm NS}=1.4$\Msun \, and a He-star (the
progenitor of the second NS) of mass $M_o$ in an orbit of semi-major axis
$A_o$. We assume that the pre-SN orbit is circular because of an earlier
mass transfer phase, during which the first NS got mildly recycled. At the
SN explosion we assume that the newborn NS is also of $M_{\rm
NS}=1.4$\Msun \, and it receives an {\em isotropic} kick of given
magnitude, $V_k$. Our assumption about the NS masses is well justified by
the measured masses (or sum of two masses) in all the observed NS/NS
binaries (\cite{T99}1999).

For a given set of values for the input parameters, $M_o$, $A_o$, and
$V_k$, we can use conservation laws of energy and angular momentum for the
system and derive expressions for the post-SN orbital semi-major axis,
$A$, and eccentricity, $e$ (see also \cite{H83}1983): 
 \begin{equation}
 A~=~\frac{\beta\,A_o}{2\beta-u_k^2\,\sin^2\theta -(u_k\cos\theta+1)^2}
 \end{equation}
 \begin{equation}
 1-e^2~=~\frac{1}{\beta^2}\left[u_k^2\,\sin^2\theta\,\cos^2\phi+
(u_k\cos\theta+1)^2\right]\,\left[2\beta-u_k^2\,\sin^2\theta
-(u_k\cos\theta+1)^2\right].
 \end{equation}
 In the above equation $u_k$ is the kick magnitude in units of the pre-SN
relative orbital velocity, $V_{\rm orb}=[G(M_{\rm NS}+M_o)/A_o]^{1/2}$,
$\beta\equiv (M_{\rm NS}+M_{\rm NS})/(M_{\rm NS}+M_o)$, and angles
$\theta$ and $\phi$ describe the direction of the kick: $\theta$ is the
polar angle from the pre-SN orbital velocity vector of the exploding
He-star and ranges from $0-\pi$ (at $\theta=0, \vec{V_k}$ and $\vec{V_{\rm
orb}}$ are aligned); $\phi$ is the azimuthal angle in the plane
perpendicular to $\vec{V_{\rm orb}}$ and ranges from $0-2\pi$ (at $\phi=0$,
the kick component in that plane points towards the first NS).

\subsection{Constraints}

Considerations of the complete set of constraints imposed on the angles
$\theta$ and $\phi$ lead to a few interesting constraints on $u_k$ and the
ratio $A/A_o$. We confirm two limits on $u_k$ found earlier by
\cite{BP95}(1995) and discuss two more limits imposed on $A/A_o$. Given an
amount of mass loss during the SN, i.e., given a value of $\beta$, the
binary gets disrupted unless
 \begin{equation} 
 u_k~<~1+\sqrt{2\beta}.
 \end{equation}
 If more than half of the total binary mass is lost ($\beta < 0.5$) then
there is a minimum kick magnitude that is required to keep the system
bound
 \begin{equation}
 u_k~>~1-\sqrt{2\beta}.
 \end{equation}
 For a given amount of mass loss and kick magnitude there is a {\em lower}
limit to the degree of orbital contraction (or expansion) that can be
achieved after the explosion
 \begin{equation}
 \frac{A}{A_o}~>~\left[2-\frac{(u_k-1)^2}{\beta}\right]^{-1}.
 \end{equation}
 Finally, if a kick is not necessary to keep the system bound ($\beta >
0.5$) but it still imparted and its magnitude is moderate, $u_k <
\sqrt{2\beta}-1$, then there is also an {\em upper} limit to the degree of
orbital contraction (or expansion) that can be
achieved after the explosion
 \begin{equation}  
 \frac{A}{A_o}~<~\left[2-\frac{(u_k+1)^2}{\beta}\right]^{-1}.
 \end{equation}
 It can be easily shown that the right part of equation (6) is smaller
than that of equation (7) for any $\beta$ and $u_k$. We note that these
limits on $A/A_o$ are imposed in addition to the two long-known limits
(\cite{F75})
 \begin{equation}
 \frac{1}{1+e} < \frac{A}{A_o} < \frac{1}{1-e}.
 \end{equation}

\subsection{Distribution Functions}

To calculate the branching probabilities for each of the three post-SN
outcomes described in \S\,2 we use a semi-analytical method based on 
Jacobian transformations of distribution functions that describe the
characteristics of the populations after the explosion.

For isotropic kicks, the distribution function of their direction, 
is given by:
 \begin{equation}
 G(\theta,\phi)~=~\frac{\sin\theta}{2}\,\frac{1}{2\pi}
 \end{equation}
 We can obtain the distribution function of $A$ and $e$ using the
Jacobian transformation of the phase space ($\theta$,$\phi$) to that of
($A$,$e$). The necessary derivatives are calculated analytically from
equations (2) and (3). It is:
 \begin{equation}
 F(A,e)~=~G(\theta,\phi)~J\left(\frac{\theta,\phi}{A,e}\right)
 \end{equation}
 and the final outcome:
 \begin{eqnarray}
 F(A,e)&=&\frac{\beta^2 e}{2\pi u_k A}\left[\beta(1-e^2)A/A_o-
\left(\frac{2\beta-\beta A_o/A -u_k^2
-1}{2}+1\right)^2\right]^{-1/2} \nonumber \\
 &&\times \left[2\beta-\beta A_o/A-\beta(1-e^2)A/A_o\right]^{-1/2}.
 \end{eqnarray}
The branching probabilities,\PCB , \PWB , \PS , are then calculated by
numerically integrating the above function over the appropriate ranges of
values of $A$ and $e$ (see Fig.\ 1) with $M_o$, $A_o$, and $V_k$ as input
parameters. 

The ranges of values for $M_o$ and $A_o$ are in principle unconstrained
unless one models the full binary evolution prior to the stage of
immediate NS/NS progenitors. However, here we are interested in obtaining
a conservative upper limit to the NS/NS coalescence rate and therefore to
the ratio (\PCB /\PS )$^{\rm max}$, so we choose the most conservative
(widest) limits to these ranges that would still lead to the formation of
coalescing NS/NS. The lower limit to $M_o$ is set by the requirement that
the He-star is massive enough to form a NS. It has been estimated to lie
between $2-3$\Msun (e.g., \cite{H85}1985) and we adopt $M_o^{\rm
min}=2$\Msun .  The upper limit so that BH formation is avoided is not
well known but simulations of the collapse of He-stars place it at $\sim
10$\Msun \,(\cite{F99a}1999). To be conservative we adopt $M_o^{\rm
max}=20$\Msun .  For a given value of $M_o$, the range of $A_o$ is
limited by the requirement that \PCB \,is {\em non-zero}. It is
interesting to note that the possible worrysome case of non-zero values
of both \PCB \, and \PS \, does not occur for {\em any} values of the
input parameters.

For a given kick magnitude we scan the complete range of $M_o$ and $A_o$
and calculate (\PCB /\PS )$^{\rm max}$. We note that the maxima occur at
$M_o\sim 4-5$\Msun \, and $A_o\sim 4-5$\Rsun \, both of which are
reasonably expected values for He-star masses and orbital sizes after
common envelope evolution. Our results for the maximum probability ratio
as a function of kick magnitude are shown in Figure 2 for a wide range of
$V_k$ from 0 to 2500\,km\,s$^{-1}$. It is evident that there is a
pronounced peak at $\sim 400$\,km\,s$^{-1}$ that gives us a unique upper
limit of (\PCB /\PS )$^{\rm max}\approx 0.27$.

\section{Empirical Coalescence Rates}

The basic method for obtaining empirical birth rates of pulsar
populations has been described in detail in \cite{N87}(1987) and other
papers (e.g., \cite{L93}1993). In short, for each observed pulsar a scale
factor is calculated based on the inverse of the ratio of the Galactic
volume throughout which the pulsar could have been detected (given the
radio pulsar surveys) over the total Galactic volume (modulo an assumed
Galactic spatial distribution of radio pulsars). This scale factor
essentially measures how many more pulsars like the one observed exist in
the Galaxy. Combined with an estimate of the lifetime of the pulsar and
summed over all observed pulsars of interest an estimate for their
Galactic birth rate can be obtained.

We apply this method (following \cite{L93}1993) to the subset of single
radio pulsars that appear to be ``similar'' to the ones detected in NS/NS
binaries. A crucial issue is how one defines this subset. Our limited
theoretical understanding of the details of the evolution of pulsar
properties (spin period $P$ and magnetic field strength $B$) does not
allow a clear definition of these ``NS/NS-like'' isolated pulsars in
terms of their observed or inferred characteristics. Instead we adopt
here a different, rather empirical way of separating this subset which
however appears to be quite promising.

A contour map of the estimated total number of pulsars in the Galaxy for
each observed pulsar in the P--B plane presented by \cite{D95}(1995)
shows quite clearly separate concentrations of pulsars in the plane. Most
of them follow a double peaked distribution with a valley (of low pulsar
number) separating the two peaks. The appearance of this valley is found
to be statistically significant at 98.37\% confidence level. At the
low-field end of the lower peak island contours include the pulsars found
in NS/NS binaries as well as a few isolated pulsars.  We choose these few
isolated pulsars (excluding the ones in globular clusters) with an
inferred magnetic field strength lower than $10^{11}$\,G to estimate
$BR_S$. Their scale factors and lifetimes are given in Table 1. We note
that for their lifetimes we have simply used their characteristic ages
defined as $\tau_c\equiv P/2\dot{P}$. Of the five pulsars only one (PSR
J2235+1506) has a position on the P-B plane that would naturally identify
it as a member of the subgroup of interest to us. The rest are chosen as
described above. To acknowledge this uncertainty related to identifying
the pulsars that originated from NS/NS progenitors we calculate our upper
limit using the estimated rate from only PSR J2235+1506 and from all
pulsars in Table 1. The range of their birth rate is $8.3\times
10^{-7}-1.8\times 10^{-5}$\,yr$^{-1}$.

So far we have ignored the possibility that a pulsar could remain
undetected if its beam does not intercept our line of
sight. Estimations of the mean beaming fraction (the fraction of
$4\pi$ steradians covered covered by the radio beam in one rotation)
have fluctuated significantly over the years (c.f.~\cite{NV83}1983;
\cite{LM88}1988; \cite{B90}1990).  Although recent work by
\cite{TM98}(1998) suggest that the mean beaming fraction may be as low
as 10\%, it is generally agreed that the beaming fraction is
period-dependent, with shorter period pulsars beaming to larger
fractions of sky. In this work, where we are dealing with pulsar
periods in the range $\sim 60-400$ ms, we shall adopt a mean beaming
fraction of $\sim 30$\%, presently the most conservative estimate for
this period interval (\cite{LM88}1988).

\section{Upper Limit}

We can now combine our results on the maximum ratio of branching
probabilities and the birth rate of ``NS/NS-like'' isolated pulsars  
to obtain an upper limit to
the NS/NS coalescence rate in the Galactic disk. From equation (1) we have
 \begin{eqnarray}
 BR_{\rm CB} & \leq &\left(\frac{P_{\rm CB}}{P_S}\right)^{\rm max}~BR_S
\nonumber \\
 & \lesssim & 7\times 10^{-7} - 1.5\times 10^{-5}\,{\rm yr}^{-1}. 
 \end{eqnarray}
Comparison of the above range with the requirements
($10^{-5}$\,yr$^{-1}$) for a detection rate of 2--3 per year with
``enhanced'' LIGO shows that only its upper end could lead to
marginally optimistic detection prospects. One possibility for an
upward revision of these numbers is a correction for a population of
pulsars fainter than the lowest sensitivity limits of pulsar
surveys. As pointed out by a number of authors (e.g.~\cite{N87}1987;
\cite{L93}1993), the scale factor analyses apply only to the
population of pulsars with luminosities above that of the faintest
source in the sample. When this luminosity limit is significantly
larger than the lower limit of the underlying population, presently
thought to be around 1\,mJy\,kpc$^2$ (\cite{L98}1998), a correction
factor must be included to account for fainter sources (see
e.g.~\cite{C95}1995). For the sample we consider here, however, this
correction factor is expected to be quite small since the faintest
pulsar (B1952+29) has a luminosity of only 1\,mJy\,kpc$^2$ ---
i.e.~very close to the lower cut-off of the pulsar luminosity
function. Alternatively, \cite{KN99}(1999) point out a related bias of
underestimating the total number of pulsars in the case of small
samples of objects. Based on detailed Monte Carlo simulations
\cite{KN99}\,estimate that the likely correction factor for samples of
5 objects (i.e.~the same size as the sample in Table 1) lies between
2 and 6.

\section{Discussion}

In the derivation of the upper limit on the NS/NS coalescence rate the
choice of the sample of isolated pulsars similar to those found in
observed NS/NS binaries is a crucial step. At present it is not obvious
how such pulsars can be identified accurately in a way other than their
position on the P--B plane. The upper end of the range covered by the
upper limit derived here is quite conservative in this respect
(corresponds to a sample of 5 single pulsars). It is plausible that we
have included one or more pulsars that actually belong to the group of
usual young, non-recycled pulsars or even that they have been recycled
but in a type of binary system that could not lead to the formation of
NS/NS binaries. In either case the upper limit derived here overestimates
the true value. Note that the ratios of branching probabilities
calculated here (with a maximum value $\lesssim$ 0.3\%) indicate that the
number of single ``NS/NS-like'' pulsars should exceed that of coalescing
binaries by {\em at least} a factor of $\sim 3$. Hence, a sample bigger
than 5--6 single pulsars would be favored and there may even be a serious
deficit of such single pulsars in the observed population
(\cite{Kb99}1999).

Another issue of concern could be the validity of equation (1), that is
the equality of the ratio of birth rates to the ratio of branching
probabilities. Given that all three subgroups (of the three post-SN
outcomes) have common progenitors, the two ratios could be different if
the evolution {\em after} the recycling of the first NS is different for
each of the subgroups. The only possibility relevant to these late
evolutionary stages is related to the characteristic velocities of the
different types. However, a quick estimate of the typical center-of-mass
velocities of NS/NS binaries and of the corresponding isolated pulsars
(based on the results of \cite{K96}1996 and \cite{T98}1998) shows that
their velocities are within factors of 2--3 of the pre-SN orbital
velocities and hence of each other. Such differences are probably too
small to make any difference in the estimated birth rates, a conclusion
also supported by the recent results of \cite{E99}(1999) who find that the
dependence of the empirical pulsar birth rate estimates are very weakly
dependent on the details of the velocity and spatial distribution of the
population.

Recently, an uncertainty in empirical rate estimates related to
small-number samples and the faint end of the pulsar luminosity function
has been pointed out and studied quantitatively in the context of NS/NS
coalescence rates (\cite{KN99}1999). This uncertainty is relevant to our
empirical estimate of $BR_S$ although it is much less significant for the
upper end of the derived range. According to \cite{KN99}(1999) for an
observed sample of five objects the upward correction is probably
$\lesssim 2$.

In determining the low edge of the allowed range of $A_o$ for a given
$M_o$ we have used the constraint that the He-star lies within its Roche
lobe even at its maximum radial extent. In this way we exclude the
possibility of the NS going into a common-envelope phase with the
He-star, which we expect will lead to the collapse of the NS to a black
hole (see e.g., \cite{F96}1996; \cite{F97}1997). The lower value of $A_o$
then depends on the radius-mass relation of evolved He-stars which for
stars less massive than $\approx 3-3.5$\Msun \, is uncertain within
factors of a few. To explore the effect of this uncertainty on our
results we have repeated the calculation of (\PCB /\PS )$^{\rm max}$
reducing the He-star radii by a (relatively high) factor of 4 in the
above mass range. In this case our estimates for the limit on the
coalescence rate increases by $\approx 2.5$ ((\PCB /\PS )$^{\rm max}$
increased to 0.7).

The method proposed here to obtain upper bounds on the coalescence
rate can also be used with wide NS/NS binaries instead of the isolated
``NS/NS-like'' pulsars. The part related to the analysis of the
orbital dynamics at the second SN can easily be modified to calculate
(\PCB /\PWB )$^{\rm max}$. We performed this calculation following the
basic procedure outlined in \S\,3, scanning the complete allowed
ranges of $M_o$ and $A_o$. Our results as a function of kick magnitude
are shown in Figure 3 and again the ratio of branching probabilities
shows a clear maximum.  The second step of the derivation would
involve an empirical estimate of the birth rate of WB based on the
observed sample. At present, this sample\footnote{An former member of
this subgroup, PSR B2303+46, has been recently shown to be a NS/WD
binary (\cite{vK99}1999).}  consists of only a few objects: the mildy
relativistic binary system J1518+4904 (\cite{nst96}1996); PSR
J1811$-$1736, a 104 ms pulsar in a highly eccentric 18.8 day orbit,
discovered by the Parkes Multibeam survey (\cite{L99}1999); and
possibly the enigmatic binary pulsar B1820$-$11 (\cite{lm89}1989). The
method described here could be used in the future when more WB are
discovered and the details pertaining to the selection effects in the
most recent surveys become available. This would then provide us with
an additional constraint on the NS/NS coalescence rate.

\acknowledgements

We would like to thank R.\ Narayan, D.\ Psaltis and T.\ Prince for useful
discussions and I.\ Stairs and R.\ Manchester for sharing with us
information about pulsars recently discovered by the Parkes Multibeam
Pulsar Survey ahead of publication. Support by the Smithsonian Institute
via a Harvard-Smithsonian Center for Astrophysics Post-doctoral Fellowship
is also acknowledged. Arecibo Observatory is operated by Cornell
University under cooperative agreement with the National Science
Foundation.

\newpage

\centerline{\psfig{figure=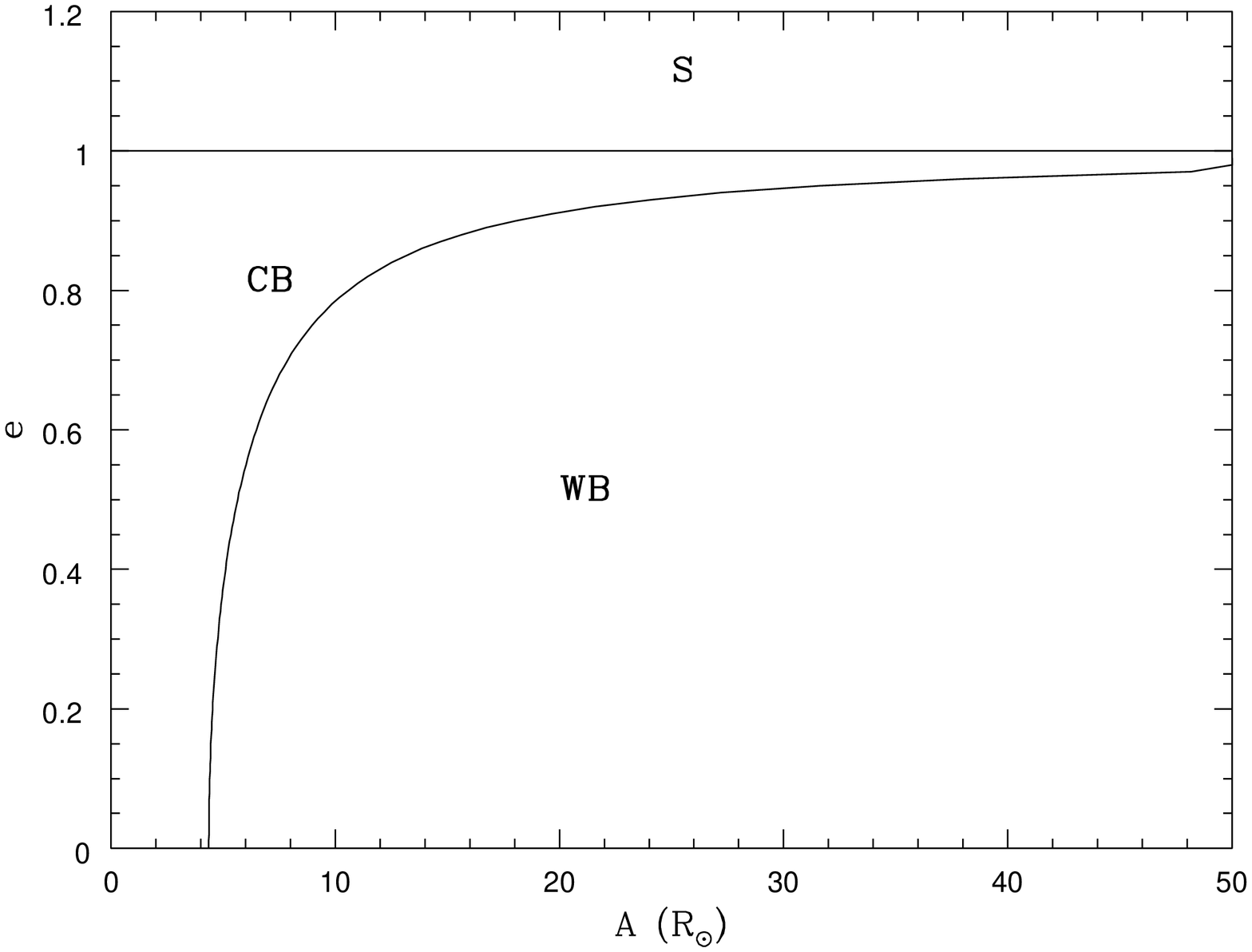,angle=-90,width=6.5in}}
\figcaption{Boundaries on the plane of post-SN orbital semi-major axis,
$A$ and post-SN eccentricity, $e$, separating (i) NS/NS binaries
coalescing within $10^{10}$\,yr (CB), (ii) NS/NS binaries too wide to
coalesce within $10^{10}$\,yr, and (iii) isolated NS freed with the
disruption of binaries when $e>1$ (S).}

\newpage

\centerline{\psfig{figure=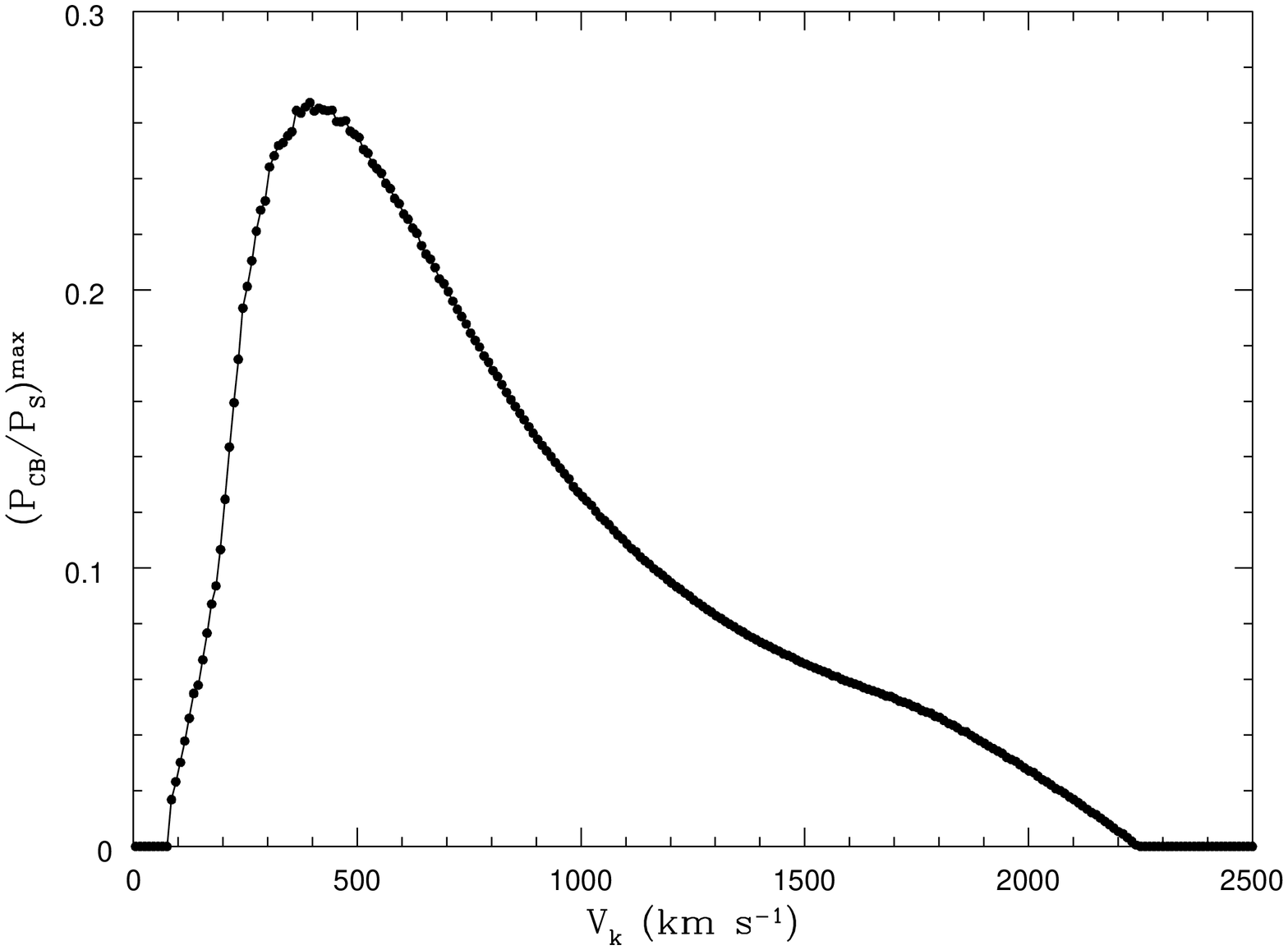,angle=-90,width=6.5in}}
\figcaption{Maximum values of the ratio of branching probabilities, \PCB
/\PS , calculated for the complete ranges of He-star masses, $M_o$, and
pre-SN orbital separations, $A_o$, plotted as a function of the
magnitude, $V_k$, of an isotropic kick imparted to the newborn NS. For
the definition of the subgroups CB and S we used the boundaries shown in
Figure 1.}

\newpage

\centerline{\psfig{figure=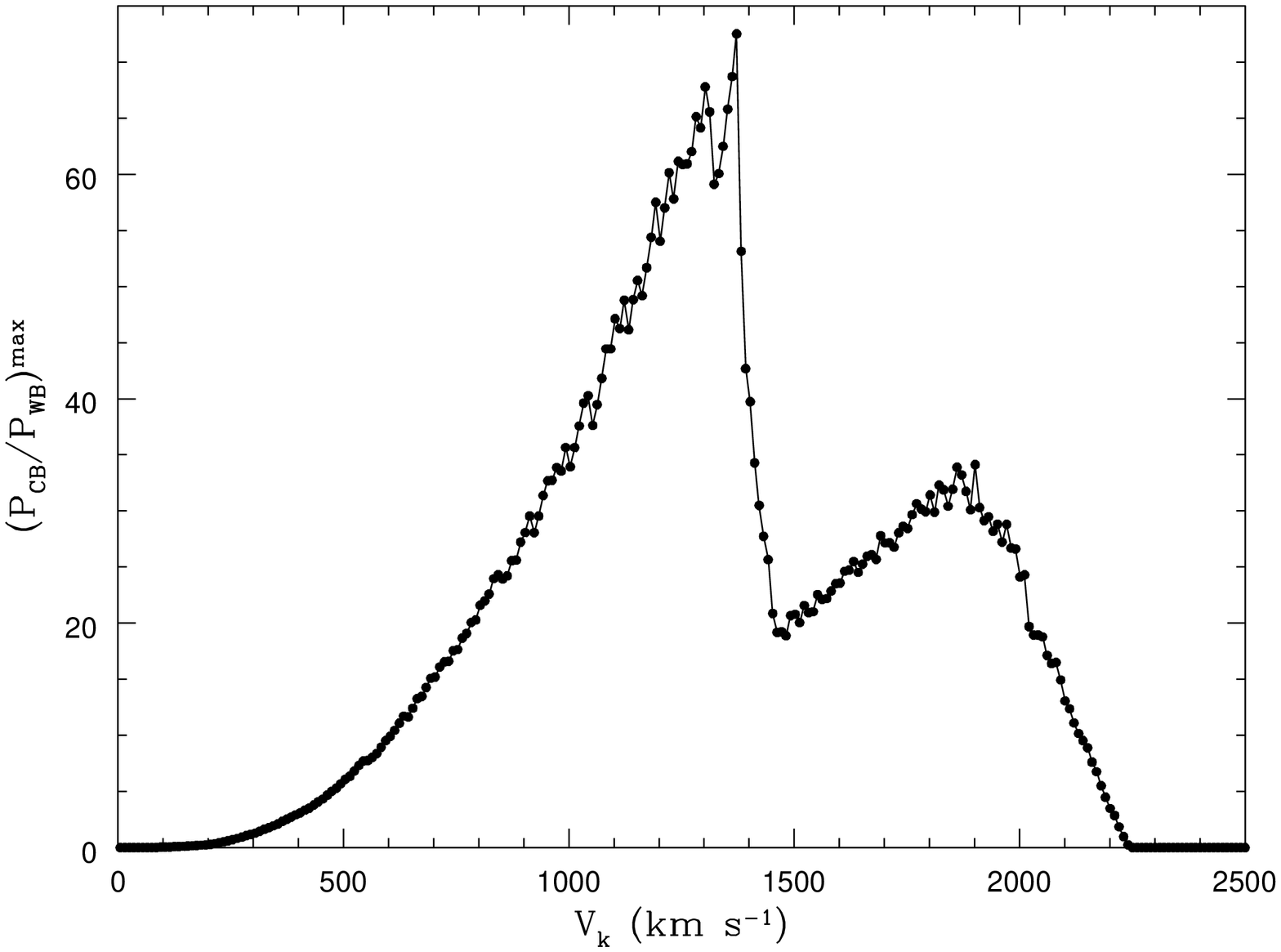,angle=-90,width=6.5in}}
\figcaption{Maximum values of the ratio of branching probabilities, \PCB
/\PWB , calculated for the complete ranges of He-star masses, $M_o$, and
pre-SN orbital separations, $A_o$, plotted as a function of the magnitude,
$V_k$, of an isotropic kick imparted to the newborn NS. For the definition
of the subgroups CB and WB we used the boundaries shown in Figure 1.}

\newpage

\begin{deluxetable}{lrrl}
\tablecolumns{4}
\tablecaption{Sample of Possible ``NS/NS-like'' Isolated Pulsars}
\tablehead{PSR Name & Scale Factor & Log($\tau_c$/yr) & $BR$
(yr$^{-1}$)}
  
\startdata

J2235+1506 & 5000 & 9.78 & 8.3$\times 10^{-7}$ \\
B1848+04 & 1000 & 9.45 & 3.55$\times 10^{-7}$ \\
B1952+29 & 15000 & 9.62 & 3.6$\times 10^{-6}$ \\
B0331+45 & 1000 & 8.76 & 1.74$\times 10^{-6}$ \\
B1804-08 & 1000 & 7.96 & 1.1$\times 10^{-5}$

\enddata

\end{deluxetable}

\end{document}